\documentclass[preprint,12pt]{elsarticle}

\usepackage{graphicx,subfigure}
\usepackage{amssymb}
\usepackage{amsmath}
\usepackage{bm}
\usepackage{lineno} 
\usepackage{natbib}
\usepackage[squaren]{SIunits}

\begin{document}

\begin{frontmatter}

\title{Modeling of Laser wakefield acceleration in Lorentz boosted frame using EM-PIC code with spectral solver}

\author[UCLAEE]{Peicheng Yu}
\ead{tpc02@ucla.edu}
\author[THUACC]{Xinlu Xu} 
\author[UCLAPH]{Viktor K. Decyk}
\author[UCLAEE]{Weiming An}
\author[IST]{Jorge Vieira}
\author[UCLAPH]{Frank S. Tsung}
\author[IST,ISCTE]{Ricardo A. Fonseca}
\author[THUACC]{Wei Lu}
\author[IST]{Luis O. Silva}
\author[UCLAEE,UCLAPH]{Warren B. Mori}

\address[UCLAEE]{Department of Electrical Engineering, University of California Los Angeles, Los Angeles, CA 90095, USA}
\address[THUACC]{Key Laboratory of Particle and Radiation Imaging of Ministry of Education, Department of Engineering Physics, Tsinghua University, Beijing 100084, China}
\address[UCLAPH]{Department of Physics and Astronomy, University of California Los Angeles, Los Angeles, CA 90095, USA}
\address[IST]{Instituto Superior T\'ecnico, Lisbon, Portugal}
\address[ISCTE]{ISCTE - Instituto Universit\'ario de Lisboa, 1649--026, Lisbon, Portugal}

\begin{abstract}
Simulating laser wakefield acceleration (LWFA) in a Lorentz boosted frame in which the plasma drifts towards the laser with $v_b$ can speedup the simulation by factors of $\gamma^2_b=(1-v^2_b/c^2)^{-1}$. In these simulations the relativistic drifting plasma inevitably induces a high frequency numerical instability that contaminates the interested physics. Various approaches have been proposed to mitigate this instability. One approach is to solve Maxwell equations in Fourier space (a spectral solver) as this has been shown to suppress the fastest growing modes of this instability in simple test problems using a simple low pass or ``ring'' or ``shell'' like filters in Fourier space. We describe the development of a fully parallelized, multi-dimensional, particle-in-cell code that uses a spectral solver to solve Maxwell's equations and that includes the ability to launch a laser using a moving antenna. This new EM-PIC code is called UPIC-EMMA and it is based on the components of the UCLA PIC framework (UPIC). We show that by using UPIC-EMMA, LWFA simulations in the boosted frames with arbitrary $\gamma_b$ can be conducted without the presence of the numerical instability. We also compare the results of a few LWFA cases for several values of $\gamma_b$, including lab frame simulations using OSIRIS, a EM-PIC code with a finite difference time domain (FDTD) Maxwell solver. These comparisons include cases in both linear and nonlinear regimes. We also investigate some issues associated with numerical dispersion in lab and boosted frame simulations and between FDTD and spectral solvers. 
\end{abstract}


\end{frontmatter}


\section{Introduction}
Laser wakefield acceleration (LWFA) offers the potential to construct compact accelerators that has a numerous potential applications including the building blocks for a next generation linear collider and being the driver for compact light sources. As a result, LWFA has attracted extensive interest since it was originally proposed \cite{TajimaDawson}, and the last ten years has seen an explosion of experimental results. Due to the strong nonlinear effects that are present in LWFA, developing predictive theoretical models is challenging \cite{LuScaling,Martins2010NatPhys}; therefore numerical simulations are critical. In particular, particle-in-cell (PIC) simulations play a very important role in LWFA research because the PIC algorithm follows the self-consistent interactions of particles through the electromagnetic fields directly calculated from the full set of Maxwell equations. Using a standard PIC code to study a 10 GeV stage in a nonlinear regime takes approximately 1 million core hours on today's computers and a 100 GeV stage would take 100 million core hours. While computing resources now exist to do a few of such simulations, it is not possible to do parameter scans in full three-dimensions. Therefore, reduced models such as combining the ponderomotive guiding center with full PIC \cite{guidingcenter} for the wake or with quasi-static PIC \cite{quasistatic1,quasistatic2} are used for parameter scans. However, while these models are very useful, they cannot model full pump depletion distances and the quasi-static approach cannot model self-injection. Another reduced model that has been recently proposed is to expand the fields in azimuthal mode numbers and truncate the expansion \cite{aztrun}. This can reduce modeling a 3D problem with low azimuthal asymmetry into the similar computational cost as using a  2D $r-z$ code. 

Recently, it was shown that by performing the simulation in an optimal Lorentz boosted frame with velocity $v_b$, the time and space scales to be resolved in a numerical simulation may be minimized \cite{Vay2007PRL,MartinsCPC2010,VayJCP2011}. The basic idea is that in the boosted frame the plasma length (the laser propagation distance) is Lorentz contracted while the plasma wake wavelength and laser pulse length are Lorentz expanded. The number of laser cycles is an invariant (assuming there is no reflected wave) so the necessary number of cells needed to resolve the laser is also an invariant while the cell size and hence time step are Lorentz expanded. The increase in time step and decrease in the plasma length lead to savings of factors of $\gamma^2_b=(1-v^2_b/c^2)^{-1}$ as compared to a lab frame simulation using the so called moving window \cite{Decker94}. Using such simulations, it has been shown that using a 1--3~$\peta\watt$ laser one could generate 10 GeV electron beam in a self-guided stage and 50 GeV in a channel guided stage \cite{MartinsCPC2010}. For these cases the savings can be larger than factors of $10^4$. However, in the boosted frame LWFA simulations noise from a numerical instability can be an issue. As discussed in \cite{Godfrey1974,Godfrey1975,Nagata2007,YuAAC,GodfreyJCP2013,XuArxiv}, the noise results from a numerical Cerenkov instability induced by the plasma drifting with relativistic speeds on the grid. According to the dispersion relation this numerical instability is attributed to the coupling between the wave-particle resonances with EM modes (including aliased modes) in the numerical system. The pattern of the instability in Fourier space can be found at the intersections of the EM dispersion relation of the solver used in the simulation algorithm, and the wave-particle resonances \cite{YuAAC,GodfreyJCP2013,XuArxiv}. 

In order to mitigate this instability, it is preferable to use an EM solver that eliminates the numerical instability at the main beam resonance. In this case, the instability occurs only at high $\vert \vec k\vert$ modes which are far away from the physics of interest. As the EM dispersion curves for most FDTD solvers inevitably bends down (i.e., supports waves with phase velocities less than the speed of light) at high $\vert \vec k\vert$, numerical instabilities at the main beam resonance are found in these solvers. However, when using a spectral solver that spatially advances the EM fields in Fourier space, the dispersion curve assures no instability pattern at the main beam resonance. In addition, the pattern at the first space aliasing beam mode is found to indeed be located at high $\vert\vec{k}\vert$ values that are far away from the interested physics. For the spectral solver the numerical Cerenkov instability is located at a predicted pattern in $\vec{k}$ space so it can be conveniently eliminated by applying simple filters directly in $\vec k$ space.

In this paper we describe the development of a fully parallelized three-dimensional electromagnetic spectral PIC code called UPIC-EMMA that was rapidly built using components of the UCLA PIC Framework (UPIC) \cite{UPIC}. We demonstrate that through the use of appropriate filters, Lorentz boosted frame simulations of LWFA at the optimum frame velocities can be carried out without limitations from the numerical Cerenkov instability. We show that a simple low pass filter with a hard cutoff at $\vert \vec k\vert$ works very well. This completely eliminates modes with $\vert \vec k\vert$ above a selected value. Meanwhile, it is not as easy to use such a filter in $\vert \vec k\vert$ space using a FDTD solver (and such solvers have instabilities at lower $\vert\vec k\vert$). 

As discussed in Ref. \cite{GodfreyJCP2013,XuArxiv}, when using the FDTD code to simulate relativistic plasma drift, an optimized time step has to be chosen to minimize the instability growth rate. While the instability growth rate is minimized, this time step does not lead to complete elimination of the instability and it can lead to further errors in numerical dispersion. Additional smoothing and filtering can help as well, but unlike when using a spectral code the instability cannot be essentially eliminated. For the spectral code, the only errors in numerical dispersion arise from the use of finite time step. Because, it is not necessary to use an optimum time step (nor does one exist), one can minimize the errors in numerical dispersion for the EM waves by choosing smaller time steps if needed. One disadvantage with the spectral code is that it is not easy to use a moving window, however for the optimum $\gamma_b$ no moving window is needed. We note that the use of a pseudo-spectral algorithm has recently been discussed and analyzed \cite{godfreyps}. This can easily be included into UPIC-EMMA if the algorithm is shown to have advantages.

We have benchmarked UPIC-EMMA by comparing the 2D and 3D simulation results of LWFA in Lorentz boosted frames with the corresponding OSIRIS \cite{OSIRIS} lab frame simulations. Good agreement is found between the OSIRIS lab frame simulations, and UPIC-EMMA boosted frame simulations, in both linear, and nonlinear regimes. We also compare UPIC-EMMA simulations for different values of $\gamma_b$ and excellent agreement is found.

The remainder of this paper is organized as follows. In section \ref{sect:numinstab} we discuss the numerical instability induced by relativistic drift. In section \ref{sect:spectralpic}, we describe the development of UPIC-EMMA, and how using the algorithms in UPIC-EMMA can eliminate the instability induced by relativistic plasma drift. In section \ref{sect:lwfaboostedframesim}, we discuss details of LWFA Lorentz boosted frame simulations using UPIC-EMMA. In section \ref{sect:lwfasim}, we benchmark UPIC-EMMA results with different $\gamma_b$ and with OSIRIS lab frame simulation. Summary is given in section \ref{sect:conclusion}.
 
\section{Numerical instability due to relativistic plasma drift}
\label{sect:numinstab}
The numerical Cerenkov instability induced by relativistic plasma drift has been extensively studied in \cite{GodfreyJCP2013,XuArxiv}. In a PIC system, when the plasma is drifting relativistically, the velocity of the drifting particles can be equal (be in resonance) to the component of the phase velocity of the main EM mode along the drift direction. In addition, its aliased modes can always be in resonance with the EM modes. The resulting wave-particle resonance leads to a violent numerical instability known as the numerical Cerenkov instability. Due to the nature of wave-particle resonances, the numerical instability occurs at the intersections of the beam resonances and EM modes determined by the Maxwell solver used in the simulation. By carefully choosing the Maxwell solver, the instability pattern can be manipulated so that mitigation can be achieved. As discussed in \cite{Godfrey1974,Nagata2007,YuAAC,XuArxiv}, when a spectral solver is used, there are no intersections of the EM modes with the main beam resonance. As a result, the instability can be found only at the aliased resonances and the fatest growing modes are the first spatial aliases. These resonances reside at high $\vert\vec{k}\vert$ in Fourier space far away from the important physics.

The instability pattern for the spectral solver can be found by investigating the corresponding dispersion relation, 
\begin{align}
&([\omega]^2-[\vec{k}]_E\cdot[\vec{k}]_B+[\vec{k}]_E[\vec{k}]_B)\vec{E}\nonumber\\
=~&-\omega^2_p \sum_{\mu,\vec{\nu}}(-1)^\mu\biggl\{\int \frac{\overleftrightarrow{S_j}(-\vec{k}')\vec{p}d\vec{p}}{\gamma\omega'-\vec{k}'\cdot\vec{p}} \biggl\{[\omega]\overleftrightarrow{S_E}(\omega',\vec{k}')\vec{E}+ \frac{\vec{p}}{\gamma}\times\{\overleftrightarrow{S_B}(\omega',\vec{k}')([\vec{k}]_E\times\vec{E})\} \biggr\}\cdot\frac{\partial f^n_0}{\partial \vec{p}}\biggr\}\nonumber
\end{align}
where $\omega$ and $\vec{k}$ are the frequency and wavenumber of the modes in the simulation system; $\vec{p}$ and $\gamma$ are the momentum and Lorentz factor of the drifting plasma; $\vec{E}$ is the electric field; $f^n_0=\delta(p_1-p_0)\delta(p_2)\delta(p_3)$ is the normalized distribution function of the plasma; $\omega_p$ is the plasma frequency; $\overleftrightarrow{S_{E}}$, $\overleftrightarrow{S_{B}}$, and $\overleftrightarrow{S_{j}}$ are the corresponding interpolation tensors for EM fields and current; $[\omega]$ and $[\vec k]_{E,B}$ are the finite difference time and space operators for the Maxwell solver used in the algorithm. And
\begin{align}\label{eqkgwg}
\omega'&=\omega+\mu\omega_g\qquad \omega_g=\frac{2\pi}{\Delta t}\qquad \mu = 0,\pm 1, \pm 2, \ldots\nonumber\\
{k}'_i&={k}_i+\nu_i{k}_{gi}\qquad {k}_{gi}=\frac{2\pi}{\Delta {x}_i}\qquad \nu_i = 0,\pm 1, \pm 2, \ldots\nonumber
\end{align}
where $\Delta t$ and $\Delta x_i$ are the time step, and grid sizes in the simulation. The sum over $\mu$ and $\vec\nu$ is attributed to the finite grid size and time step used in the simulation. Specific expressions of $[\omega]$ and $[\vec k]$ can be found in Appendix A of Ref. \cite{XuArxiv}.

\begin{figure}[b]
\begin{center}
\includegraphics[width=1.04\textwidth]{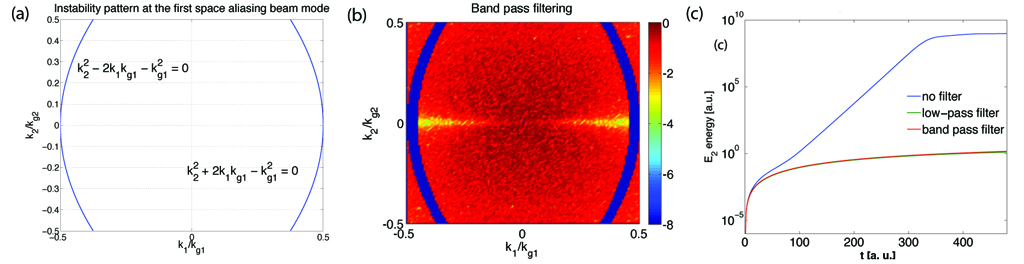}
\caption{(a) shows the analytical expression for the $\mu=0$, $\nu_1=\pm 1$ mode of numerical Cerenkov instability for the 2D spectral solver in $(k_1,k_2)$ plot; (b) shows the ``ring-shaped'' band-pass filter applied in the 2D spectral solver; and (c) shows the $E_2$ energy evolutions for various simulation setups.}
\label{fig:bandfilter}
\end{center}
\end{figure}

Since the instability pattern is found near the intersections of the EM modes and beam resonance \cite{XuArxiv}, we can obtain a simple analytical expression for the instability pattern in the limit $\Delta t\rightarrow 0$ (which leads to $[\omega]=\omega$). Under this assumption, the equation for the EM dispersion curves in the spectral solver is
\begin{align}
\omega^2\approx k^2_1+k^2_2\nonumber
\end{align}
And the equations for the beam resonances are
\begin{align}
\omega+\mu\omega_g=\beta(k_1+ \nu_1 k_{g1})\nonumber
\end{align}
where $\beta \equiv {v}/{c}$. Defining $\xi \equiv \beta\nu_1 k_{g1}-\mu\omega_g$, we can obtain the expressions for the intersections as
\begin{align} 
(1-\beta^2)k^2_1+k^2_2 - 2\beta \xi k_1 - \xi^2=0
\end{align}
Note that there are no solutions for $\mu=\nu_1=0$ for the spectral solver. The lowest order terms for the instability pattern are at the $\mu=0$ and $ \nu_1=\pm 1$ resonances. In the limit of interest $\beta\rightarrow 1$, we obtain
\begin{align}\label{eq:inst11}
k^2_2\mp 2k_1k_{g1}-k^2_{g1}=0
\end{align}
In figure \ref{fig:bandfilter} (a) we plot Eq. (\ref{eq:inst11}) for $\mu = 0$ and $\nu_1=\pm 1$. Note the ``ring'' pattern of the instability, which crosses the $k_2=0$ axis near the point $(\pm k_{g1}/2, 0)$; therefore this mode is located at high $\vert\vec{k}\vert$ values which are far away from the region of interesting physics. Therefore, the numerical Cerenkov instability can be effectively eliminated if the fatest growing modes ($\mu=0, \nu_1=\pm 1$) are suppressed in the Maxwell solver.

In the simulations, we identify the unstable modes in Fourier space using the approximate expression Eq. (\ref{eq:inst11}). We then apply filters with specific masks which multiply the undesired modes by zero. In figure \ref{fig:bandfilter} (b) we plot the  ``ring-shaped'' band-pass filter used in some of the two-dimensional simulations for testing the instability mitigation. We put ``ring'' in quotes because it is not true ring but rather a range between two  parabolas. We also used a low pass filter with a hard cut-off. We filled the simulation box with neutral plasma drifting relativistically at $\gamma=14000$ in the $x_1$ direction, and ran cases without a filter, with the ``ring'' filter, and the low pass filter with a hard cutoff. As seen from figure \ref{fig:bandfilter} (c), these filters efficiently suppresses the instability modes at $\mu=0, \nu_1=\pm 1$ in $E_2$ \cite{XuArxiv}. Therefore, the mitigation of the instability using band-pass filters shows the flexibility and efficiency of a spectral solver in being able to pinpoint the suppression of the unphysical modes in PIC simulations while leaving the modes near the interesting physics completely unaffected. 

\section{EM-PIC code with spectral solver}
\label{sect:spectralpic}
As mentioned in the section \ref{sect:numinstab}, an EM-PIC code with a spectral solver has superior properties in suppressing the numerical Cerenkov instability induced by a relativistic plasma drift. They also have superior properties with respect to numerical dispersion errors and noise. In the following, we will briefly explain the algorithm of a spectral EM-PIC code, as well as discuss the challenges in optimizing the performances of such a code.

Spectral PIC codes have a long history \cite{DawsonRMP}. However, despite their advantages in better accuracy and less noise, they are not currently as widely used because they use global field solvers which do not scale as well on parallel computers, and implementing boundary conditions is not as straight forward. A spectral EM-PIC code has the same basic flow chart as a finite-difference-time-domain (FDTD) PIC code. In a spectral EM-PIC code both the charge and current are deposited on the mesh from the particles; the forces exerted on the particles are interpolated from the mesh points, and particles are advanced using the Lorentz forces. The main difference between the spectral PIC code and FDTD PIC code is the solver used to advance the electromagnetic field and that all field quantities, including the charge and current densities, are defined at the same locations on a cell (no Yee mesh \cite{Yee} is needed). In a spectral PIC code the charge and current are directly deposited, and a strict charge conserving current deposit is not needed because Gauss's law is solved at each time step using the charge density. This gives the longitudinal part of the electric field. The longitudinal component of the magnetic field is set to zero at each time step. Faraday's law and Ampere's law are used to advance the transverse electric and magnetic fields forward in time. Note that because Gauss's law is solved for directly at each time step, a charge conserving current deposit or Boris correction to the longitudinal component of the electric field is not required to maintain that Gauss's law is satisfied. The equation for the longitudinal component of electric field $\vec E_L$ becomes:
\begin{align}
\vec{E}_L(\vec{k})=-4\pi\rho(\vec{k})\frac{i\vec{k}}{k^2}
\end{align}
and the transverse electric field and magnetic field are leap-frogged forward in time using Faraday's and Ampere's law:
\begin{align}
\frac{\partial \vec{E}_T(\vec{k})}{\partial t}&=ic\vec{k}\times\vec{B}(\vec{k})-4\pi\vec{j}_T(\vec{k})\qquad \frac{\partial \vec{B}(\vec{k})}{\partial t} = -ic\vec{k}\times\vec{E}_T(\vec{k})
\end{align} 
where the transverse component of the current is:
\begin{align}
\vec{j}_T&=\vec{j}-\frac{\vec{k}(\vec{k}\cdot\vec{j})}{k^2}
\end{align} 
We also multiply $\rho(\vec k)$ and $\vec J(\vec k)$ by a shape function $S(\vec k)=\exp(-\vert k\vert^2a^2/2)$ where $a$ is the particle size. The fields are also multiplied by this shape function then interpolated to the particles \cite{DawsonRMP}.

In addition, just as in a FDTD code, the particle positions and velocities (and correspondingly the charge and current densities) are defined at half integer values in time with respect to each other. If positions are defined at whole time steps and velocities (momentum) at half integer values, then the longitudinal and transverse components of the electric field are defined at whole time steps (when particle positions are defined) and the magnetic field is defined at half-integer values. Once the fields are transformed back from $\vec k$ space to real space then the particles can be pushed. The particle push is identical to that of a FDTD except for the interpolation of the forces because all field quantities are defined at the same locations in a cell.

There are no dispersion errors for light waves due to the grid, however there are errors from the time step. This is a significant advantage of the spectral solver, whereas a FDTD code describes the $[k]_i$ operator to $\mathcal{O}(k_i\Delta x_i)^3$, the spectral code has no errors in the $[\vec k]$ operator. Both a spectral and a FDTD code effectively truncate the highest $\vert k_i\vert$ to $\pi/\Delta x_i$. In addition, when including time step errors, the numerical dispersion of a spectral PIC code is superluminal, while that of the FDTD code is subluminal. A pseudo-spectral algorithm which also removes the time step errors has recently been described \cite{godfreyps}. As we discuss elsewhere in this paper, the more accurate and superluminal aspect of the EM dispersion relation provided by the spectral solver (together with the simple filters) is crucial for eliminating the fastest growing modes of numerical Cerenkov instability. This ensures no non-physical interaction between waves and particles in the first Brillouin zone for the spectral PIC code. The corresponding Courant condition in 2D and 3D are (for the square and cubic cells) \cite{DawsonRMP}:
\begin{align}\label{eq:spectralcl}
\Delta t_{2D} =\frac{2}{\sqrt{2}\pi c} \qquad \Delta t_{3D} = \frac{2}{\sqrt{3}\pi c}
\end{align}

A spectral PIC code is also distinguished from a FDTD code in the way it is parallelized. For the field solver, the simulation box is usually partitioned in one dimension in 2D, and two dimensions in 3D, so that each processor holds global information in the dimension to be transformed. As a result, a parallel spectral PIC code requires a fast parallel transpose routine to accomplish efficient FFT in multi-dimensions. The nature of all-to-all communications in the FFT routines makes it challenging for spectral PIC code to scale to large number of processors \cite{Spectralparallel}. In many cases the decomposition for the particles is the same as that for the fields although this does not have to be the case.

We have developed a multi-dimensional EM-PIC code using a spectral field solver called UPIC-EMMA. This code was rapidly put together using components provided by the UPIC Framework, a PIC framework with spectral solvers developed at UCLA \cite{UPIC}. UPIC-EMMA is fully relativistic and fully parallelized. Inherited from the UPIC Framework, UPIC-EMMA is coded in layers for convenient extension with different programming styles. The lowest layers are written in Fortran77 for high performance. They can be easily extended to many other languages. On top of this layer exists a library of Fortran90 wrapper functions which hide the complexity of the Fortran77 layer and that provide simpler arguments which enables strict type checking. The code separates the physics procedures from the communication, and utilizes the Message-Passing Interface (MPI) for parallel processing. In addition, a multi-tasking library was implemented to enable mixed multi-tasking and MPI messaging, where multi-tasking is used on a multiple CPU shared memory node, and message-passing is used between such nodes. UPIC-EMMA also features 3D load balancing where the fields and particles use different partitions.

\section{LWFA Simulations in the Lorentz boosted frame}
\label{sect:lwfaboostedframesim}

In section \ref{sect:numinstab} we described general issues regarding numerical instability that arises when a plasma drifts near the speed of light. In this section we describe some details regarding issues specific to modeling LWFA in a Lorentz boosted frame. We describe issues related to numerical dispersion in the lab frame, in the boosted frame, and in transforming from the boosted frame back to the lab frame for comparison. We also discuss the moving antenna and interactions between the laser and the drifting plasma boundary.

\subsection{Numerical dispersion errors for the laser}
\begin{figure}[p]
\begin{center}
\includegraphics[width=1.04\textwidth]{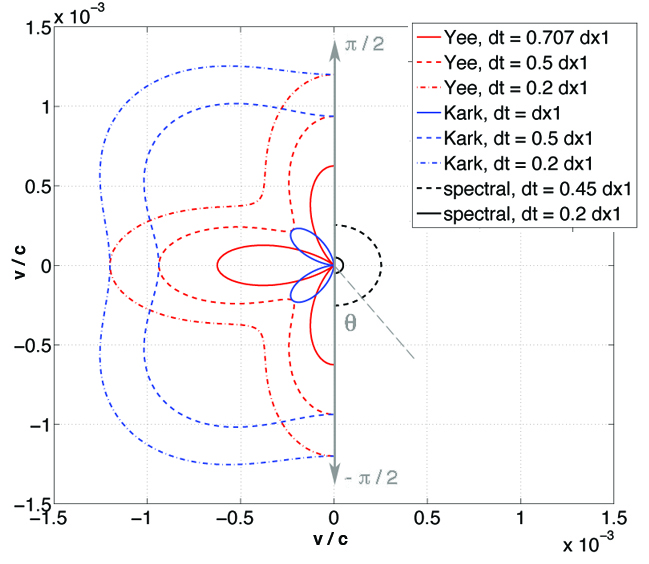}
\caption{The plot shows the errors in the group velocity defined as $v_g-1$ of the 2D EM dispersion relation for various cases. Defining $\theta=0$ to be the laser propagating direction, this plot shows the propagation angle in $(-\pi/2,\pi/2)$. If the error $(v_{g}-1)$ is larger than zero, its corresponding point is in the right side of the vertical axis, and vice versa. The group velocity is calculated for the $k_0=1.0$ mode while we are using $k_0\Delta x_1=k_0\Delta x_2=0.1$ for the calculation.}
\label{fig:groupvelocityerror}
\end{center}
\end{figure}

As mentioned in the introduction, one of the first obstacles in modeling LWFA in a boosted frame is to mitigate the numerical Cerenkov instability. For the FDTD PIC code \cite{GodfreyJCP2013,XuArxiv} which uses a combination of a Yee solver together with the momentum conserving field interpolation scheme, it is useful to choose the optimal time step $\Delta t \approx \Delta x_1/2$, where $\hat{1}$-direction is the plasma drifting direction, to minimize the numerical Cerenkov instability growth rate. The need to use this time step eliminates the flexibility in tuning the time step to minimize numerical dispersion errors for the laser. In figure \ref{fig:groupvelocityerror} we present the error in group velocity of an EM wave on a grid in 2D (we let $\Delta x_1=\Delta x_2$). Note that for the Yee, and Karkkainen solvers \cite{kark} which were discussed in Ref. \cite{GodfreyJCP2013,XuArxiv}, the most accurate dispersion relation occurs at their Courant Limit, but not the corresponding optimal time step at $\Delta t \approx \Delta x_1/2$ (for momentum conserving field interpolation). On the other hand, for a spectral PIC code the instability mitigation does not rely on the relation of grid sizes and time step. In particular, the EM dispersion relation can be made arbitrarily accurate by reducing the time step [see figure \ref{fig:reflection} (a)]. Therefore, in general when simulating relativistically drifting plasma, a spectral PIC code can provide more accuracy and flexibility over the FDTD PIC code with respect to numerical dispersion in the simulated frame. Note that recently in Ref. \cite{godfreyps} a pseudo-spectral algorithm is described which can further improve the accuracy.

\subsection{Lorentz transform of boosted frame data}
\label{sect:ltlwfa}
While numerical dispersion errors exist when using a finite size grid in vacuum, here we show that when modeling the LWFA in the Lorentz boosted frame, these errors in the boosted frame are not necessarily an issue when the results are transformed back to the lab frame. While the value for $\gamma_b$ in the boosted frame can be arbitrary, the speed up is generally larger as $\gamma_b$ is increased. However, choosing $\gamma_b \approx \gamma_w$, where $\gamma_w$ is the phase velocity of the wake, is generally optimum because in this frame the plasma length and the laser pulse length are nearly matched. When the laser and plasma frequency are comparable each is resolved similarly, i.e., there is no over-resolution of either the laser wavelength or plasma wavelength. In the boosted frame, the length of the plasma contracts by $\gamma_b$, the electron and ion mass are both $\gamma_b$ times heavier, the plasma density is $\gamma_b$ times larger, and the corresponding plasma frequency is a Lorentz invariant. As for the laser, there is a $\gamma_b(1+\beta_b)$ stretch to the pulse length, while the Rayleigh length contracts by $\gamma_b$. Therefore, while the pulse waist does not change, the effective spot size at the rear of the pulse increases by a factor of $\gamma^2_b(1+\beta_b)$. Hence for sufficiently large $\gamma_b$ an antenna is needed to launch the laser from the laser pulse waist that is moving backwards. The antenna is usually placed at the plasma boundary (see section \ref{sect:antenna} for details).

In the lab frame simulation, a moving window which only models the region of interest around the laser is often used to reduce the simulation box size. Implementation of a moving window is challenging in a spectral PIC code due to the non-local nature of the field solver which necessitates knowledge of boundary condition at both of the moving boundaries. However, the relative range of $x_1$ and $t$ contracts when Lorentz transforming the data of interest from lab frame to boosted frame. If $\gamma_b$ is appropriately chosen, in this frame the length of the plasma column is of the same order as the laser pulse length. As a result, for $\gamma_b\sim\gamma_w$ it is feasible to conduct the boosted frame simulation without the moving window.

\begin{figure}[p]
\begin{center}
\includegraphics[width=1.04\textwidth]{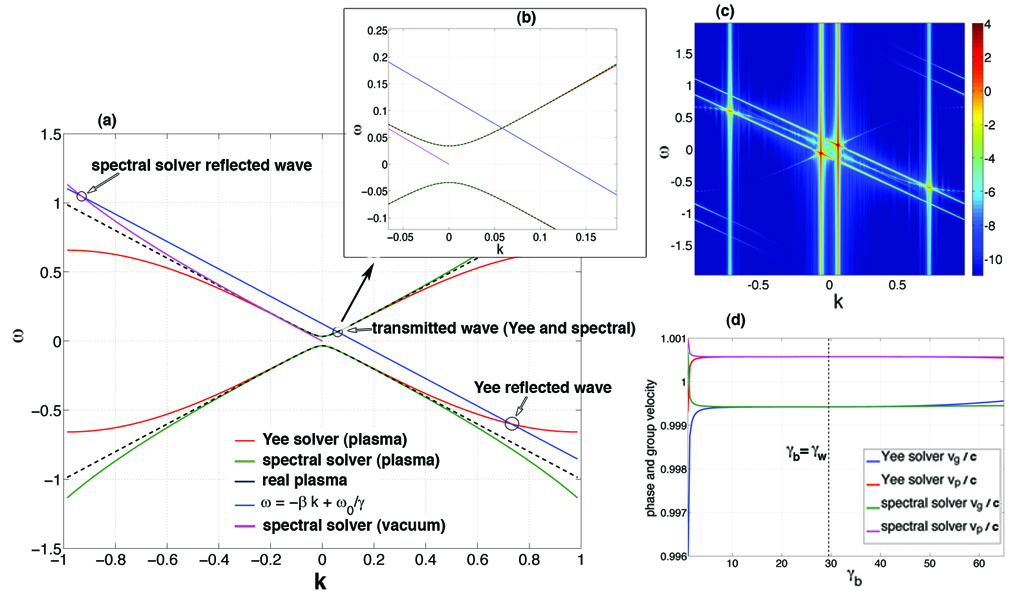}
\caption{(a) shows the intersections of the line $\omega = -\beta_bk+\frac{\omega_0}{\gamma_b}$ and various EM dispersion curves, while in (b) we magnified the region near the origin; (c) shows an example of the $E_3$ spectrum of a 1D LWFA boosted frame simulation with the Yee solver. The hot spots in (c) show where the transmitted and reflected waves are, and agrees with the prediction in (a). (d) shows the dependence of the transformed phase and group velocity of the EM waves in the plasma with $\gamma_b$. The phase and group velocity converges quickly as $\gamma_b$ increases from 1.}
\label{fig:reflection}
\end{center}
\end{figure}

In LWFA lab frame simulations, an EM wave with frequency $\omega_0$ is incident on a stationary plasma slab. This leads to a reflected and transmitted waves, each having the incident frequency. Their wave numbers are determined from the dispersion relation in vacuum (reflected wave), and in plasma (transmitted wave). In a simulation the same physics occurs except the EM wave now satisfies the numerical dispersion relation in vacuum and plasma. In the boosted frame there is still a reflected and transmitted waves, except in this case the incident wave, reflected wave, and transmitted wave each have different frequencies. Furthermore, numerical issues can lead to some subtle effects. An effective method to identify the frequencies of the reflected and transmitted waves is to use an $(\omega,k)$ diagram, which was previously used in studying the radiation generated from ionization fronts \cite{wkdiag}. At the plasma boundary $z_0=-v_bt$, the phase of each wave $\phi=kz-\omega t=-(kv_b+\omega)t$ must be the same, otherwise the continuity of fields cannot be satisfied at every instant in time. This leads to
\begin{align}
k_iv_b+w_i=-k_rv_b+\omega_r=k_tv_b+\omega_t
\end{align}
where $i$, $r$, $t$ correspond to incident, reflected, and transmitted waves respectively. For example, if $v_b=0$ then $\omega_i=\omega_r=\omega_t$. If the incident and reflected waves obey the vacuum dispersion relation $\omega=ck$ then 
\begin{align}
\omega_r=\frac{1+\beta_b\omega_i}{1-\beta_b}
\end{align}
which can also be obtained from a double Lorentz transformation. In a Lorentz boosted frame the plasma is drifting but $\omega_i=\omega_0$ is Lorentz transformed to $\omega'_i$ and we want $\omega'_r$ and $\omega'_t$ [where the $(')$ sign refers to the boosted frame variables]. In this frame
\begin{align}\label{eq:straightline}
\omega'+k'v_b=\omega'_0+k'_0v_b
\end{align}
where $\omega'$ can be either the reflected or transmitted waves. The constant $\omega'_0+k'_0v_b$ is obtained by Lorentz transforming $\omega_0$ and $k_0$ into the boosted frame: $\omega'_0=\gamma_0(\omega_0-k_0v_b)$ and $k'_0=\gamma_0(k_0-v_b\omega_0/c^2)$. Therefore, $\omega'+k'v_b=\omega_0/\gamma_b$ regardless of the relationship between $\omega_0$ and $k_0$. In a real system $\omega_0=k_0c$ although numerical errors in the dispersion relation do not alter the constant $\omega_0/\gamma_b$. Therefore, the reflected and transmitted waves must fall along the line $\omega'=-k'v_b+\omega_0/\gamma_b$ in $(\omega',k')$ space (here we are ignoring the aliasing modes). In addition, they must also fall on the dispersion curves for light in a plasma \cite{XuArxiv}
\begin{align}\label{eq:plcurve}
[\omega]^2=[k]^2+\frac{\omega'^2_p}{\gamma_b}S^2\frac{[\omega]-[k]v_b}{\omega-kv_b}
\end{align}
 or in vacuum
 \begin{align}\label{eq:vacurve}
[\omega]^2=[k]^2
\end{align}
on the grid where we assume $S=S_{j3}=S_{E3}=S_{B2}$ in Eq. (19) of Ref. \cite{XuArxiv}, and $\omega'^2_p/\gamma_b=4\pi e^2n'_0/m_e$ is Lorentz invariant where $n_0$ is the lab frame density, $e$ is the electron charge, and $m_e$ is the electron rest mass. The reflected and transmitted wave lie at the intersection between Eq. (\ref{eq:straightline}), and Eq. (\ref{eq:plcurve}) in plasma or Eq. (\ref{eq:vacurve}) in vacuum. This is shown in figure \ref{fig:reflection} (a) for a case where $\Delta t\approx 0.5 \Delta x_1$, $\omega_0/\omega_p\approx 30$, and $\gamma_b=8.0$. The line $\omega'=-v_bk'+\omega_0/\gamma_b$ and the dispersion curve for a real plasma (black dashed lines), a FDTD Yee solver (red lines), and a spectral solver (green lines for inside the plasma; magenta lines for in vacuum) are shown. The vacuum dispersion relation is plotted for the spectral solver in the upper left quadrant for the reasons given in the next paragraph. In figure \ref{fig:reflection} (b), we have expanded the region in $(\omega,k)$ space near the origin to illustrate the frequency and direction of the transmitted wave which does not depend strongly on the solvers used. When $\gamma_b=\gamma_w=\omega_0/\omega_p$ then $\omega'_t=\omega_p=\omega'_p/\sqrt{\gamma_b}$ and $k'_t=0$. If $\gamma_b>\gamma_w$ then $\omega'_0$ would be negative and the phase velocity and group velocity of the transmitted wave would be negative; however, since $\vert v'_{gt}\vert<\vert v_b\vert$ the transmitted wave would still be in the plasma.

Figure \ref{fig:reflection} (a) also illustrates that numerical errors to the dispersion relation effect the location of the reflected wave. In a real system where $\omega'=ck'$ in vacuum and $\omega'^2=\omega'^2_p/\gamma^2_b+c^2k'^2$ in the plasma, then the reflected wave would occur where $\omega'=-\beta_bk'+\omega_0/\gamma_b$ intersects the vacuum curve, i.e. at $\omega'=\omega_b\gamma_b(1+\beta_b)\sim 2\omega\gamma_b$, which is larger than the largest $\omega'$ in the fundamental Brillouin zone. However, for the numerical dispersion curves shown in figure \ref{fig:reflection} (a), the reflected wave resides at the intersection with the plasma dispersion relation in the lower right quadrant for the FDTD solver or with the vacuum dispersion relation in the upper left quadrant for the spectral solver. For the FDTD case, the reflected wave has a negative phase and group velocity. However, the group velocity is less than $v_b$ so the reflected wave propagates backwards while staying inside the plasma. For the spectral solver the group velocity is slightly larger than the speed of light so it resides outside the plasma. The predicted locations of the transmitted and reflected waves are confirmed in an OSIRIS (FDTD) simulation. This is seen in figure \ref{fig:reflection} (c), in which the $\omega'$ and $k'$ spectrum is plotted from a simulation for parameters identical to those used to generate the theoretical plot in figure \ref{fig:reflection} (a). Strong signals are seen at the predicted locations. For cases of interest the reflection coefficient is small [the reflected signal is significantly smaller than the transmitted signal in figure \ref{fig:reflection} (c)] so the unphysical mode is not energetically important, and it does not complicate the physics. 

We have also investigated the invariance of transforming results back to the lab frame based on the numerical dispersion relations. We note that solving Maxwell's equations on a grid using discrete time steps is not strictly Lorentz invariant. For example, the group velocity of light in vacuum for a spectral solver is greater than the speed of light, and it depends on $\omega\Delta t$. Nevertheless, when carrying out LWFA (or other) simulations in a boosted frame, the results are transformed back to the lab frame using the Lorentz transformations. This is assumed to be reasonable if one is looking at modes which are properly resolved.

As noted earlier, when $\gamma_b$ is chosen near $\gamma_w$ there is a balance between the laser pulse length and the plasma length. In addition, for $\gamma_b\approx\gamma_w$ the transmitted wave $k'\sim 0$, and errors in the boosted frame due to the finite cell size are minimized. In figure \ref{fig:reflection} (d) we show that when $v'_\phi$ and $v'_g$ for the transmitted wave are Lorentz transformed back to the lab frame using the velocity addition formulas, 
\begin{align}\label{eq:phasegroupv}
\beta_\phi=\frac{\beta'_\phi+\beta_b}{1+\beta'_\phi \beta_b}\qquad \beta_g=\frac{\beta'_g+\beta_b}{1+\beta'_g \beta_b}
\end{align}
where $\beta_\phi$ and $\beta_g$ are the phase and group velocity normalized to $c$, that the numerical errors are nearly absent for sufficiently large $\gamma_b$. 
The values for   $v'_\phi$ and $v'_g$ are calculated from the linear dispersion relation, where $\omega \Delta t=0.5k\Delta x_1$ is given and $k\Delta x_1=0.2$ is from the dispersion relation in the lab frame. In the boosted frame $\Delta t'=\gamma_b(1+\beta_b)\Delta t$ and $\Delta x'_1=\gamma_b(1+\beta_b)\Delta x_1$. According to the plot, for $\gamma_b=1$ there are clear numerical errors; however, for $\gamma_b\ge 5$, the numerical errors are minimized. 

These results can be understood as follows. In a Lorentz boosted frame where $\beta_b=\beta_w\equiv (1-\gamma^{-2}_w)^{1/2}$, the group velocity $\beta'_g\rightarrow 0$, while the phase velocity $\beta'_\phi\rightarrow \infty$ in the numerical system, which when substituted back to Eq. (\ref{eq:phasegroupv}) leads to
\begin{align}
\beta_\phi=1/\beta_w\qquad \beta_g=\beta_w
\end{align}
which are the accurate values for a continuous system. In addition, writing $\beta=\overline\beta+\delta_\beta$, and $\beta'=\overline\beta'+\delta_\beta'$, where $\overline\beta$ and $\overline\beta'$ corresponds to the correct values in the lab and boosted frame, and defining
\begin{align}
\overline\beta_{\phi,g}=\frac{\overline\beta'_{\phi,g}+\beta_b}{1+\overline\beta'_{\phi,g} \beta_b}
\end{align}
we can obtain the expressions of the error $\delta_\beta\equiv \beta_{\phi,g}-\overline\beta_{\phi,g}$ as
\begin{align}\label{eq:verrorgammab}
\delta_\beta=\frac{1}{(1+\overline\beta'\beta_b/\delta_\beta'+\beta_b)(1+\overline\beta'\beta_b/\delta_\beta')\gamma^2_b}
\end{align}
Note the $\gamma^2_b$ in the denominator indicates that when $\gamma_b$ is sufficiently large, the errors in velocity when transformed back to the lab frame will be small for any waves. We also note that the arguments going from Eq. (\ref{eq:phasegroupv}) to Eq. (\ref{eq:verrorgammab}) hold for any velocity including those of the particles. This indicates that if we choose the $\gamma_b$ large enough that we would obtain more accurate results compared to a simulation done in the lab frame (with typical cell sizes and time steps). An area of future work is to quantify the differences more accurately.

\subsection{Moving antenna}
\label{sect:antenna}
\begin{figure}[p]
\begin{center}
\includegraphics[width=1.04\textwidth]{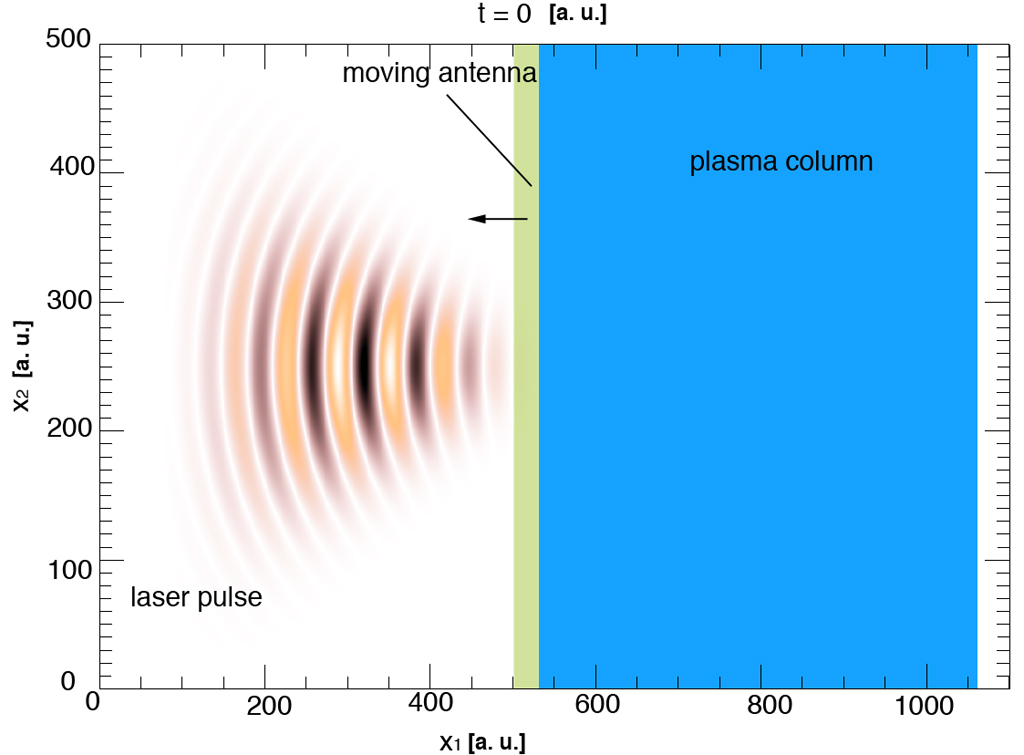}
\caption{UPIC-EMMA simulation setup for LWFA boosted frame simulation. The blue block is the plasma column; the green slice is the moving antenna at $t=0$. The laser is launched via the moving antenna (moving together with the plasma column boundary at $v=-\beta_b$) by initializing the appropriate curent in the green slice which has a typical width of $\lambda/2$. The laser is likewise plotted for $t=0$. Note when the laser is launched via the antenna, only the area within the antenna is initialized. }
\label{fig:simsetup}
\end{center}
\end{figure}

As discussed in \cite{MartinsCPC2010} and \cite{VayArxiv2011}, the effective spot size of the laser increases by a factor of $\gamma_b^2(1+\beta_b)$ because the Rayleigh length of the laser contracts by $\gamma_b$ and the pulse length expands by $\gamma_b (1+\beta_b)$. To prevent the need for using a simulation box with transverse size $\sim\gamma^2_b$ times that in needed in the lab frame, we utilize a thin slice of grids at the plasma boundary (where the laser beam waist resides) as an antenna to drive the laser pulse into the plasma. The antenna is moving together with the plasma boundary [see figure \ref{fig:simsetup}]. 

The EM field in the moving antenna as a function of $\vec{x}$ and time $t$ can be derived as follows (see also \cite{MartinsCPC2010} and \cite{VayArxiv2011}). For instance, for a laser linearly polarized in the $\hat{2}$ direction, the expression for the electric field $E_2(\vec{x},t)$ of a Gaussian pulse in the lab frame can be expressed as:
\begin{align}
E_2(x_1,x_2,x_3,t)=\frac{E_0W_0}{W(x_1)}&{\exp\biggl[-\frac{x^2_2+x^2_3}{W^2(x_1)}\biggr]}{\exp\biggl[-\frac{2(x_1-ct)^2}{\sigma^2_s}\biggr]}\nonumber\\
&{\exp\biggl[ikx_1+ik\frac{x^2_2+x^2_3}{2R(x_1)}-i\textrm{arctan}\frac{x_1}{x_R}\biggr]}{\exp(-i\omega t)}\nonumber
\end{align}
with
\begin{align}
&{W(x_1)=W_0 \sqrt{1+\frac{x^2_1}{x^2_R}}}\qquad {R(x_1)=x_1\biggl(1+\frac{x^2_R}{x^2_1}\biggr)}\qquad {x_R=\frac{\pi W^2_0}{\lambda}}\nonumber
\end{align}
where $E_0$ is the amplitude, $W_0$ is laser pulse waist, $\sigma_s$ is the laser pulse length, $\omega$ and $k$ are the laser frequency and wavenumber, and $x_R$ is the laser Rayleigh length. A similar expression holds for the magnetic field $B_3(\vec{x},t)$. After Lorentz transforming, we can readily obtain the new expression of the laser pulse in the boosted frame
\begin{align}
\label{eq:antennaE}E'_2(\eta,x'_2,x'_3,t')=~&\frac{E'_0W_0}{W'(\eta)}{\exp\biggl[-\frac{x'^2_2+x'^2_3}{W'^2(\eta)}\biggr]}{\exp\biggl[-\frac{2[\eta-(1+\beta)ct']^2}{\sigma'^2_s}\biggr]}\nonumber\\
&{\exp\biggl[ik'\eta+ik'\frac{x'^2_2+x'^2_3}{2R'(\eta)}-i\textrm{arctan}\frac{\eta}{x'_R}\biggr]}{\exp(-i\omega' t')}
\end{align}
where
\begin{align}
\label{eqs112}\eta=x'_1+\beta_b ct'\qquad
\sigma'_s=\gamma_b(1+\beta_b)\sigma_s\qquad E'_0=\frac{E_0}{\gamma_b(1+\beta_b)}\\
k'=\frac{k}{\gamma_b(1+\beta_b)}\qquad \omega'=\frac{\omega}{\gamma_b}\qquad x'_R=\frac{x_R}{\gamma_b}\\
W'(\eta)=W_0\sqrt{1+\frac{\eta^2}{x'^2_R}}\qquad R'(\eta)=\frac{\eta}{1+\beta_b}\biggl(1+\frac{x'^2_R}{\eta^2}\biggr)
\end{align}
We have neglected the fact that $E_2$ may not equal $B_3$ when solving for the fields on a grid, i.e., $\omega \neq kc$. In the spectral code, the transverse and longitudinal components of the fields are solved for separately. Therefore, on the antenna we set $\rho=0$ so there are no longitudinal fields on it. When launching a laser from the antenna, we assign current (in the direction of the laser polarization direction) at every point inside the antenna such that $\vec{E}$ has the desired form and polarization. The other components and the magnetic field follow naturally from the Maxwell field solver. The antenna has a finite width of around $\lambda/2$ where $\lambda$ is the wavelength of the laser in vacuum to eliminate any backward propagating signal. The current for generating the laser is added after the current is deposited for all the particles in the system is finished.

\begin{figure}[htbp]
\begin{center}
\includegraphics[width=1.04\textwidth]{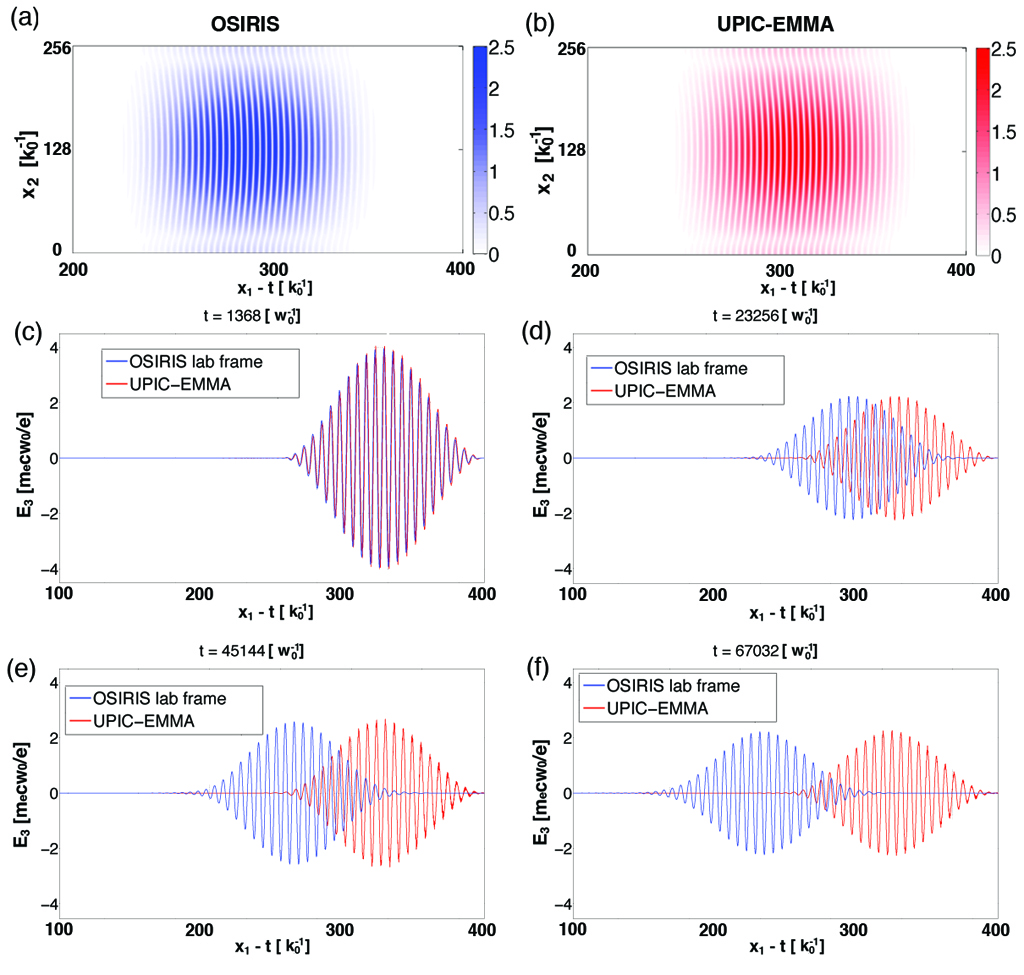}
\caption{(a) is the 2D plot of the laser (polarized in $x_3$ direction) $E_3$ field at $t=13680~\omega^{-1}_0$, and (b) shows the laser $E_3$ field transformed back from the boosted frame data. (c)--(f) shows the comparison of on-axis $E_3$ field between OSIRIS data and UPIC-EMMA data at various time points. $x_1-t$ is the coordinates moving together with the moving window.}
\label{fig:movant}
\end{center}
\end{figure}

The moving antenna implemented in UPIC-EMMA is benchmarked by transforming the data back to the lab frame and then comparing it to data from an OSIRIS lab frame simulation. In the OSIRIS lab frame run, the laser propagates in the $x_1$ direction together with the moving window; as in the UPIC-EMMA run, the laser is launched from a moving antenna. In the UPIC-EMMA simulation $\gamma_b=14$ is used. Periodic boundary condition are used for transverse directions in both cases. The transformed UPIC-EMMA boosted frame data (to the lab frame) are plotted together with the lab frame OSIRIS data in figure \ref{fig:movant}. Good agreement is found between the two cases. Note the shift in the laser wave packet between the OSIRIS data and UPIC-EMMA data. We verified that the shift was attributed to the difference in group velocity between the Yee solver and spectral solver (transformed back to lab frame). 

\subsection{Filters}
Earlier the mode numbers of the fastest growing modes of the numerical Cerenkov instability in the spectral solver were identified as Eq. (\ref{eq:inst11}). Based on this equation, we use filters that eliminate a range of $\vec k$'s centered around this parabola. Specifically, we muliply all modes by either 1 or 0. Those modes multiplied by 0 are those in the range:
\begin{align}
k^2_2=\pm 2k_{g1}(k_1+\Delta k_1)
\end{align}
in 2D, and 
\begin{align}
k^2_2+k^2_3=\pm 2k_{g1}(k_1+\Delta k_1)
\end{align}
in 3D. $\Delta k_1$ is usually chosen to be
\begin{align}
0.9\times\frac{k_{g1}}{2} < \Delta k_1 < 1.02\times\frac{k_{g1}}{2}
\end{align}

\section{LWFA simulations with UPIC-EMMA}
\label{sect:lwfasim}
We next present simulation results using UPIC-EMMA to model LWFA in a boosted frame. Two-dimensional simulations in the linear and nonlinear regimes are presented for two different choices of $\gamma_b$ and the results are compared to OSIRIS simulation results in the lab frame (the UPIC-EMMA results are transformed back to the lab frame). We also present 3D results from UPIC-EMMA including comparison with OSIRIS lab frame simulations.  For the linear cases $a_0=0.1$, while for the nonlinear cases $a_0$=3.0 or 4.0.  In both the OSIRIS and UPIC-EMMA simulations the time step is chosen near the Courant limit. Precise values for the simulations parameters are  shown in tables \ref{tab:simpara1} and \ref{tab:simpara2}. 

\begin{figure}[htbp]
\begin{center}
\includegraphics[width=1.0\textwidth]{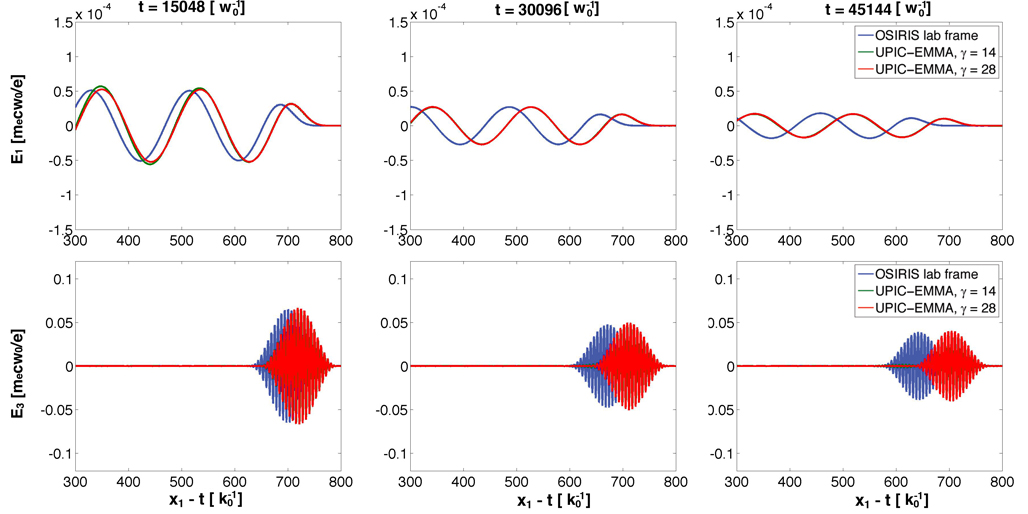}
\caption{Comparison of the on-axis $E_1$ and $E_3$ between OSIRIS lab frame simulation, and UPIC-EMMA boosted frame simulation ($\gamma=14,28$) at various time steps, for $a_0=0.1$. $x_1-t$ is the coordinates moving together with the moving window. }
\label{fig:ad1}
\end{center}
\end{figure}

In figure \ref{fig:ad1}, results from the $a_0$=0.1 case are shown. In the top row, the wakefield $E_1$ is shown at various lab frame times for an OSIRIS lab frame simulations (blue) and for two UPIC-EMMA simulations where $\gamma_b$=14 (red) and $\gamma_b$=28 (green) respectively. This figure shows that for early times the three curves are inseparable while for later times the results from OSIRIS lag behind. At later times the UPIC-EMMA results for the different $\gamma_b$ remain inseparable. There is no evidence in these plots of any noise in the wake and laser fields due to the numerical Cerenkov instability. The fact that the OSIRIS lab frame result slips backwards is due to the numerical dispersion error in $v_g$ that was discussed earlier. In the bottom row of figure \ref{fig:ad1}, the laser field ($E_3$) is plotted at the same times. The same colors are used to show the results from the three simulations. The slippage of the OSIRIS lab frame curve is also seen in the laser field. 

We next show results for a more nonlinear case where $a_0$=3.0. As before, there is a lab frame OSIRIS simulation and two UPIC-EMMA boosted frame simulations with $\gamma_b$=14  and $\gamma_b$=28. The same colors as in  figure \ref{fig:ad1} are used to distinguish the data from these three simulations. We plot the accelerating field in the upper row and the laser field in the lower row at four various lab frame times (different times than used in figure \ref{fig:ad1}). Similarly to the linear $a_0=0.1$ case the wakefields from the three simulations are inseparable at early times while for later times the OSIRIS results slip behind. While the agreement between the two boosted frame simulations is still good, it is not as good as for the previous case. The differences in the laser field are small for larger values of $x_1-t$ (at the head of the laser) and there are differences at later times. 

\begin{figure}[p]
\begin{center}
\includegraphics[width=1.04\textwidth]{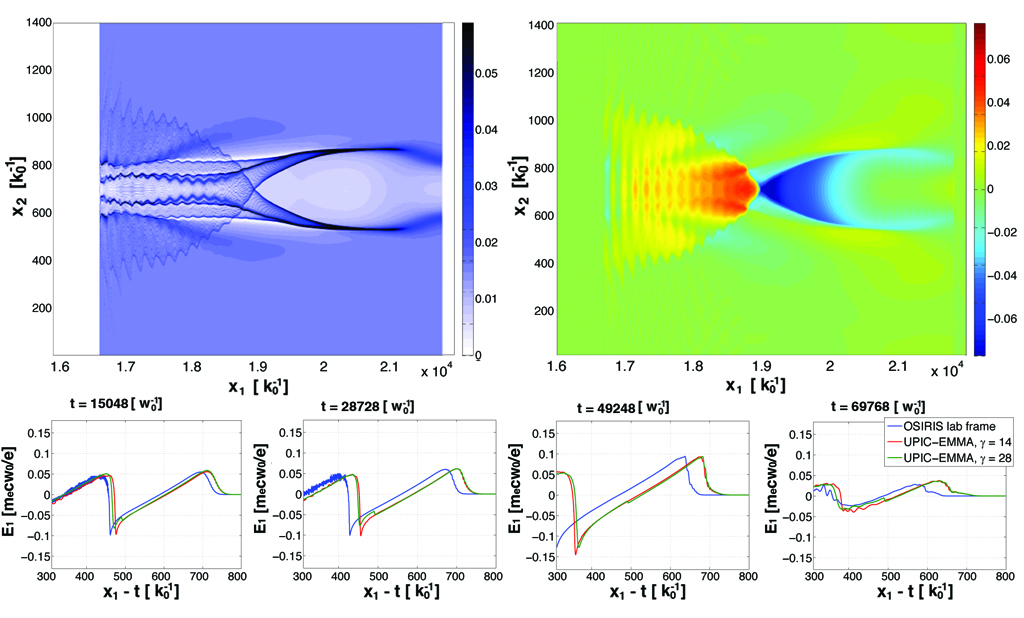}
\caption{UPIC-EMMA boosted frame simulation ($\gamma=14,28$) for $a_0=4.0$. First row shows the 2D plots of plasma electron density (left), and the corresponding $E_1$ for $t'=6180~\omega^{-1}_0$ in the boosted frame ($\gamma=14$). The second row shows the on-axis $E_1$ comparison between OSIRIS lab frame, and UPIC-EMMA boosted frame simulation ($\gamma=14,28$). $x_1-t$ is the coordinates moving together with the moving window. }
\label{fig:a4e1}
\end{center}
\end{figure}

\begin{figure}[p]
\begin{center}
\includegraphics[width=1.04\textwidth]{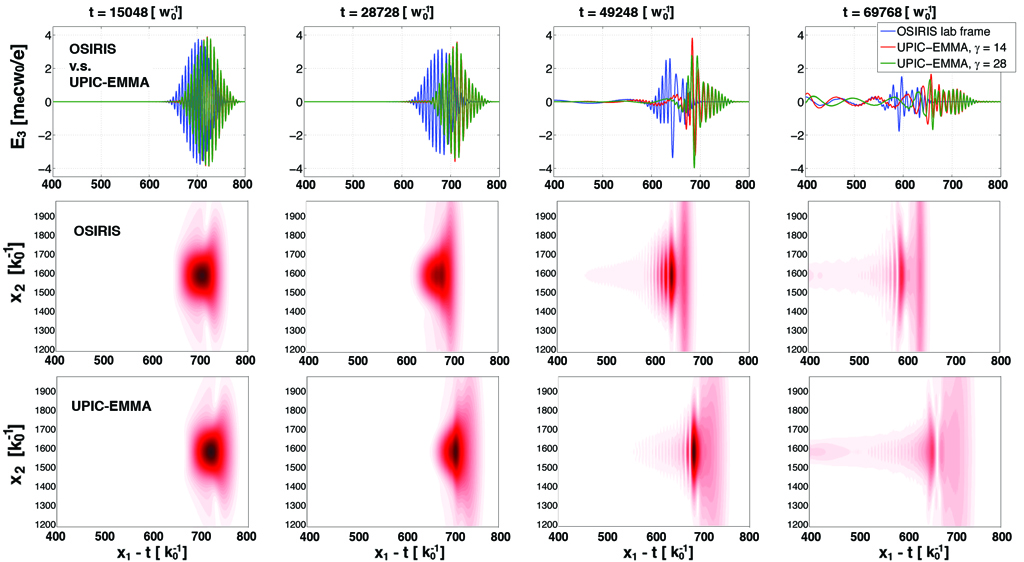}
\caption{Comparison of the $E_3$ field between OSIRIS lab frame simulation, and UPIC-EMMA boosted frame simulation ($\gamma=14,28$) at various time steps, for $a_0=4.0$. The first row shows on-axis $E_3$ comparison between OSIRIS lab frame, and UPIC-EMMA boosted frame ($\gamma=14,28$). The second and third rows show the 2D comparison between the OSIRIS lab frame results and the transformed data from UPIC-EMMA boosted frame ($\gamma=14$). $x_1-t$ is the coordinates moving together with the moving window. }
\label{fig:a4e3}
\end{center}
\end{figure}

Next in figures \ref{fig:a4e1} and  \ref{fig:a4e3}, we present results from a case where the laser amplitude is increased to  $a_0=4.0$.   In the top row of figure \ref{fig:a4e1}, the plasma density and wakefield in the Lorentz boosted frame with $\gamma_b=14$ are plotted at x$t'=6180\omega_0^{-1}$. There is no evidence of the numerical Cerenkov instability. Only a small region of the simulation box, including where the instability is most robust, is plotted. In the lower row of figure \ref{fig:a4e1} we also plot in the lab frame the wakefields obtained in these three simulations. As in the two previous cases, good agreement is found in the wakefield amplitude. There is slippage of the wakefield in the OSIRIS simulation and small differences between the two boosted frame simulations.

It is worth noting that, in the case where $\gamma_b=28$ (green), the wakefield around the spike looks different compared with that of $\gamma=14$ (red) and OSIRIS lab frame simulation (blue). The flattened part in the $\gamma=28$ curve indicates more self-trapped charge is loaded in the wakefield. We believe these differences may be due to different statistics since each macro-particle represents much higher charge at higher $\gamma_b$. In the LWFA lab frame simulation we usually choose the longitudinal cell sizes (along the laser propagation direction) to be a fraction of $c/\omega_0$, and the perpendicular cell size to be a fraction of $c/\omega_p$, where $\omega_0$ and $\omega_p$ are the frequency of the laser and plasma. Meanwhile, in the Lorentz boosted frame the perpendicular cell size remains the same, while the longitudinal cell size increases by $\gamma_b(1+\beta_b)$ due to the stretch of the laser wavelength. At the same time, the plasma column contracts by $\gamma_b$ thus the plasma density increases by $\gamma_b$. As a result, if we keep the number of particle per cell to be the same as in the lab frame, factors of $(1+\beta_b)\gamma^2_b$ savings can be achieved in the boosted frame simulation. Howver, in this case the increase in longitudinal cell sizes and plasma density causes one macro-particle to represent $(1+\beta_b)\gamma^2_b$ more charge then in the lab frame. For example, if we use a cell with longitudinal size of $0.2c/\omega_0$, perpendicular sizes of $0.2c/\omega_p$, and 8 particles per cell, for a plasma density $\sim 10^{18}~\centi\meter^{-3}$,  each macro-particle represent $\sim 3.6\times 10^3$ real electrons, which corresponds to $\sim 0.6$ fC of charge. If $\gamma_b\sim 30$, then one macro-particle in the boosted frame corresponds to $\sim 1.2$ pC. This shows that modeling self-trapping in Lorentz boosted frames at large $\gamma_b$ will require future work including identifying where the self-trapped particles come from and loading more particles per cell in these regions.

Figure \ref{fig:a4e3} shows the comparison of the laser $E_3$ fields for the three $a_0$=4.0 cases. In the top row we show line outs of the laser at four different propagation distances (times). The OSIRIS lab frame curve slips backwards. As in the other cases, the boosted frame curves line up at the front of the laser and as in the other nonlinear case differences in the curves are seen in the back of the laser. In addition,  for the $\gamma_b=14$ case we transform not only the on-axis data, but also the off-axis data in order to compare the 2D laser profile between OSIRIS lab frame run and UPIC-EMMA boosted frame run. The OSIRIS lab frame data is shown in the middle row and the UPIC-EMMA data in the bottom row. Only a part of the the simulation box is shown. Good agreement is found in how the laser pump depletes between the two runs and in how the shape evolves. The slippage of the OSIRIS simulation results is seen.

\begin{figure}[p]
\begin{center}
\includegraphics[width=1.04\textwidth]{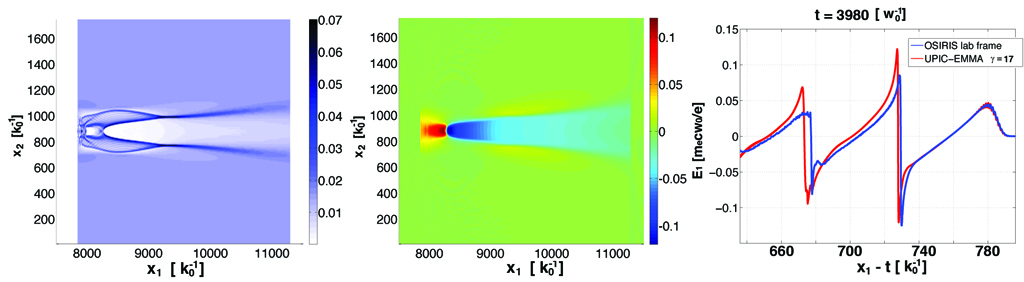}
\caption{Results from 3D UPIC-EMMA boosted frame simulation ($\gamma=17$). (a) and (b) present 2D cross section plot of the plasma electron density, and $E_1$ at $t' = 15335~\omega^{-1}_0$, while (c) shows the on-axis $E_1$ comparison at $t=3980~\omega^{-1}_0$ in the lab frame. $x_1-t$ is the coordinates moving together with the moving window.}
\label{fig:lu3d}
\end{center}
\end{figure}

Last, to illustrate that UPIC-EMMA is fully working in three-dimensions, we present the 3D results of UPC-EMMA using the simulation parameters in table \ref{tab:simpara2} and $a_0$=4.0. These parameters are similar to those in Ref. \cite{LuScaling}. In figure \ref{fig:lu3d} (a) and (b) we present 2D slices in the boosted frame of the  plasma density and wakefield at the center of the box in the $\hat{3}$-direction at $t'=15335\omega_0^{-1}$. As in the 2D cases, no noise from the numerical Cerenkov instability is evident. In figure \ref{fig:lu3d} (c) , the wakefield at $t=3980~\omega^{-1}_0$ in the lab frame  from the OSIRIS lab frame simulation (blue) and UPIC-EMMA boosted frame simulation (red) are shown. The curves agree well but not perfectly.  Note that in (c), there is no slippage because we are showing the result at a time where little slippage has occurred. Future work will involve understanding these differences for these nonlinear cases. 

\begin{table}[t]
\centering
\begin{tabular}{lr}
\hline\hline
Plasma density $n_0$& $1.148\times 10^{-3} n_0\gamma_b$\\
Laser & \\
\quad pulse length $\tau$ & $ 70.64k^{-1}_0\gamma_b(1+\beta_b)$\\
\quad puse waist $W$ & $117.81k^{-1}_0$\\
\quad polarization & $\hat 3$-direction\\
\hline
Lab frame simulation $(\gamma_b=1)$&\\
\quad grid size $(\Delta x_{1},\Delta x_{2})$ & $(0.2k^{-1}_0, 2.746k^{-1}_0)$\\
\quad time step $\Delta t$ & $0.199\omega^{-1}_0$\\
\quad number of grid (moving window) & $4000\times 512$ \\
\quad particle shape & quadratic\\
\hline
2D boosted frame simulation&\\
\quad grid size $\Delta x_{1,2}$ & $0.0982k^{-1}_0\gamma_b(1+\beta_b)$\\
\quad time step $\Delta t$ & $0.0221\omega^{-1}_0\gamma_b(1+\beta_b)$\\
\quad number of grid $(\gamma_b=14)$& 16384$\times$512 \\
\quad number of grid $(\gamma_b=28)$& 8192$\times$256\\
\quad particle shape & quadratic\\
\hline\hline
\end{tabular}
\caption{Simulation parameters for the 2D simulations, with $a_0=0.1, 3.0, 4.0$ (related to figure \ref{fig:ad1}, \ref{fig:a3}, \ref{fig:a4e1}, and \ref{fig:a4e3}). The laser frequency $\omega_0$ and laser wave number $k_0$ are used to normalize simulation parameters, and $n_0=m_e\omega^2_0/(4\pi e^2)$.}
\label{tab:simpara1}
\end{table}

\begin{table}[t]
\centering
\begin{tabular}{lr}
\hline\hline
Plasma density $n_0$& $8.611\times 10^{-4} n_0\gamma_b$\\
Laser & \\
\quad pulse length $\tau$ & $ 70.64k^{-1}_0\gamma_b(1+\beta_b)$\\
\quad puse waist $W$ & $117.81k^{-1}_0$\\
\quad polarization & circular\\
\hline
Lab frame simulation $(\gamma_b=1)$&\\
\quad grid size $(\Delta x_{1},\Delta x_{2},\Delta x_{3})$ & $(0.2k^{-1}_0, 3.40k^{-1}_0,3.40k^{-1}_0)$\\
\quad time step $\Delta t$ & $0.199\omega^{-1}_0$\\
\quad number of grid (moving window) & $4000\times 512 \times 512$ \\
\quad particle shape & quadratic\\
\hline
3D boosted frame simulation&\\
\quad grid size $\Delta x_{1,2,3}$ & $0.2k^{-1}_0\gamma_b(1+\beta_b)$\\
\quad time step $\Delta t$ & $0.04\omega^{-1}_0\gamma_b(1+\beta_b)$\\
\quad number of grid $(\gamma_b=17)$ & $4096\times 256 \times 256$ \\
\quad particle shape & quadratic\\
\hline\hline
\end{tabular}
\caption{Simulation parameters for the 3D simulations (related to figure \ref{fig:lu3d}). The laser frequency $\omega_0$ and laser wave number $k_0$ are used to normalize simulation parameters, and $n_0=m_e\omega^2_0/(4\pi e^2)$.}
\label{tab:simpara2}
\end{table}

\begin{figure}[p]
\begin{center}
\includegraphics[width=1.05\textwidth]{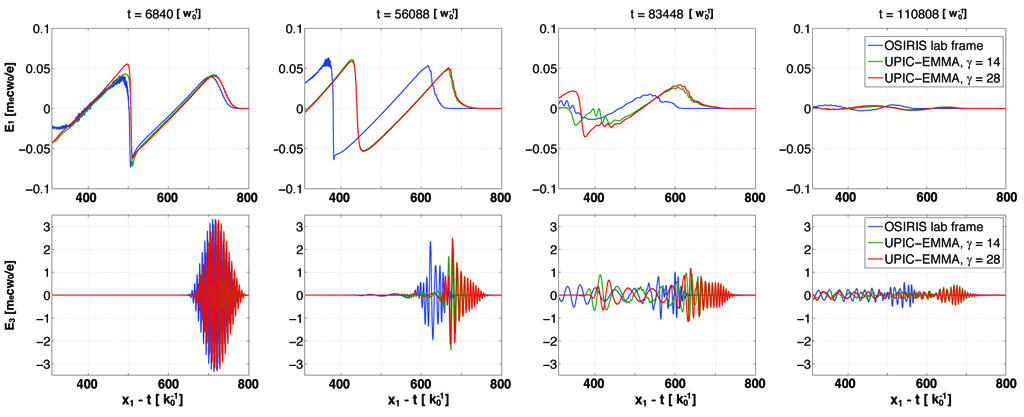}
\caption{Comparison of the on-axis $E_1$ and $E_3$ between OSIRIS lab frame simulation, and UPIC-EMMA boosted frame simulation ($\gamma=14,28$) at various time steps, for $a_0=3.0$. $x_1-t$ is the coordinates moving together with the moving window. }
\label{fig:a3}
\end{center}
\end{figure}

\section{Summary}
\label{sect:conclusion}

In this paper, we described the rapid development of a new three dimensional PIC code that can be used to model laser wakefield acceleration in the Lorentz boosted frames. In such simulations a plasma is drifting at relativistic speeds towards the laser, which leads to the numerical Cerenkov instability. The growth rates and unstable mode numbers of the numerical Cerenkov instability depends on the type of Maxwell field solver used. The new code, called UPIC-EMMA, uses a spectral field solver, and is fully parallelized. It is built using the components of the UPIC Framework, which is a set of modules for building parallelized PIC codes with FFT based (spectral) solvers.  The use of a spectral solver in which the fields are solved for in Fourier space allows for more convenient mitigation of the numerical Cerenkov instability. The phase velocity of light in vacuum and in a plasma is always greater than the speed of light for a spectral solver. In such cases, the fastest growing modes of the numerical Cerenkov instability are due to the lowest order aliased beam mode and they reside at large values of $\vert\vec k\vert$. These modes  can be easily filtered out using a ``hard'' low pass or ``shell'' filters, thereby eliminating the fast growing modes of the instability.

We presented examples of LWFA boosted frame simulations using UPIC-EMMA. Several different values of the laser amplitude were simulated ranging from a very linear regime to a nonlinear regime. For the cases shown there was no evidence of the numerical instability and good agreement was found between OSIRIS lab frame and UPIC-EMMA boosted frame simulations. The comparison showed that the wake and laser from OSIRIS lab frame simulation slipped behind the results from the boosted frame simulations as expected from numerical dispersion errors. We showed that the dispersion errors become smaller when results are transformed back the lab frame. 

The results indicate that the use of a spectral code may be attractive for carrying out high fidelity LWFA simulations in boosted frames at high $\gamma_b$. Future work will involve using pseudo-spectral methods, studying how self-trapping occurs in boosted frames where each macro-particle contains significant charge,  studying and understanding the differences between boosted frame and lab frame simulations for these and more nonlinear regimes, and using a moving cathode to simultaneously launch trailing beams into laser driven wakes or to study beam driven wakes. 

This work was supported by DOE awards DE-FC02-07ER41500, DE-SC0008491, DE-FG02-92ER40727, and DE-SC0008316, and by NSF grants NSF PHY-0904039 and NSF PHY-0936266, and by NSFC Grant 11175102, thousand young talents program, and by FCT (Portugal), grant EXPL/FIS-PLA/0834/1012, and by the European Research Council (ERC-2010-AdG Grant 267841). Simulations were carried out on the UCLA Hoffman 2 Cluster, and Dawson 2 Cluster.


\end{document}